\documentclass{emulateapj}

\pdfoutput=1
\usepackage{epstopdf}
\usepackage{natbib}
\usepackage{amsmath}

\newcommand{\gri}{\protect\hbox{$gri$} }

\newcommand{\griz}{\protect\hbox{$griz$} }

\newcommand{\gione}{\protect\hbox{$gi[3.6]$} }

\newcommand{\about}{$\sim\!\!$~}

\newcommand{\sqdeg}{\protect\hbox{deg$^2$} }

\newcommand{\err}[2]{\ensuremath{^{+#1}_{-#2}}}
\newcommand{\msun}{M$_{\sun}$}
\newcommand{\munit}{$h_{70}^{-1}$~M$_{\sun}$}
\def\lsim{\hbox{\rlap{\raise 0.425ex\hbox{$<$}}\lower 0.65ex\hbox{$\sim$}}}
\def\gsim{\hbox{\rlap{\raise 0.425ex\hbox{$>$}}\lower 0.65ex\hbox{$\sim$}}}
\def\arcmin{\hbox{$^\prime$}}
\def\arcsec{\hbox{$^{\prime\prime}$}}
\def\arcdeg{\mbox{$^\circ$}}

\newcommand{\cluster}{SPT-CL~J2106-5844}

\newcommand{\masssz}{\ensuremath{(1.06 \pm 0.23) \times 10^{15}}}

\newcommand{\massyxf}{\ensuremath{(0.93 \pm 0.19) \times 10^{15}}}
 
\newcommand{\masstxf}{\ensuremath{(0.92 \pm 0.37) \times 10^{15}}}

\newcommand{\massdyn}{\ensuremath{1.3^{+1.4}_{-0.5} \times 10^{15}}}
\newcommand{\massdynpass}{\ensuremath{1.4^{+1.7}_{-0.8} \times 10^{15}}}
\newcommand{\massbest}{\ensuremath{(1.27 \pm 0.21) \times 10^{15}}}
\newcommand{\zx}{$z = 1.18 \pm 0.03$}
\newcommand{\zxnoe}{$z = 1.18$}

\newcommand{\zs}{$z = 1.131^{+0.002}_{-0.003}$}
\newcommand{\zsnoe}{$z = 1.132$}
\newcommand{\zspas}{$z = 1.132^{+0.002}_{-0.003}$}
\newcommand{\xtemp}{$kT = 11.0^{+2.6}_{-1.9}$~keV}
\newcommand{\xlum}{$L_{X} ({\rm 0.5 - 2.0~keV}) = (13.9 \pm 1.0) \times 10^{44}$~erg~s$^{-1}$}
\newcommand{\uk}{\mu \mathrm{K}}
\newcommand{\disp}{$\sigma_v = 1230^{+270}_{-180}$~km~s$^{-1}$}
\newcommand{\disppas}{$\sigma_v = 1270^{+320}_{-210}$~km~s$^{-1}$}

\newcommand{\disppast}{\ensuremath{1270^{+310}_{-220}}}
\newcommand{\xtempt}{\ensuremath{11.0^{+2.6}_{-1.9}}}
\newcommand{\masssztf}{\ensuremath{1.06 \pm 0.23}}
\newcommand{\massyxttf}{\ensuremath{1.85 \pm 0.40}}
\newcommand{\masstxttf}{\ensuremath{1.83 \pm 0.76}} 
\newcommand{\massdynpasstf}{\ensuremath{1.4^{+1.7}_{-0.8}}}
\newcommand{\massbesttf}{\ensuremath{1.27 \pm 0.21}}

\shorttitle{Discovery and Cosmological Implications of \cluster}
\shortauthors{Foley et~al.}

\def\CfA{1}
\def\Clay{2}
\def\Munich{3}
\def\MIT{4}
\def\ExcellenceCluster{5}
\def\McGill{6}
\def\Harvard{7}
\def\Cardiff{8}
\def\UChicago{9}
\def\NCSA{10}
\def\KICPChicago{11}
\def\EFIChicago{12}
\def\PhysicsUChicago{13}
\def\NASA{14}
\def\AAUChicago{15}
\def\PUC{16}
\def\Illinois{17}
\def\PSU{18}
\def\Berkeley{19}
\def\UFlorida{20}
\def\Colorado{21}
\def\Davis{22}
\def\LBNL{23}
\def\Michigan{24}
\def\MPE{25}
\def\CaseWestern{26}
\def\Caltech{27}
\def\STScI{28}
\def\Yale{29}

\begin{document}

 \title{Discovery and Cosmological Implications of \cluster, the Most
 Massive Known Cluster at \lowercase{$z > 1$}}


\author{
R.~J.~Foley\altaffilmark{\CfA,\Clay},
K.~Andersson\altaffilmark{\Munich,\MIT},
G.~Bazin\altaffilmark{\Munich,\ExcellenceCluster},
T.~de~Haan\altaffilmark{\McGill},
J.~Ruel\altaffilmark{\Harvard},  
P.~A.~R.~Ade\altaffilmark{\Cardiff},
K.~A.~Aird\altaffilmark{\UChicago},
R.~Armstrong\altaffilmark{\NCSA},
M.~L.~N.~Ashby\altaffilmark{\Harvard},
M.~Bautz\altaffilmark{\MIT},
B.~A.~Benson\altaffilmark{\KICPChicago,\EFIChicago},
L.~E.~Bleem\altaffilmark{\KICPChicago,\PhysicsUChicago},
M.~Bonamente\altaffilmark{\NASA},
M.~Brodwin\altaffilmark{\CfA},
J.~E.~Carlstrom\altaffilmark{\KICPChicago,\EFIChicago,\PhysicsUChicago,\AAUChicago},
C.~L.~Chang\altaffilmark{\KICPChicago,\EFIChicago},
A.~Clocchiatti\altaffilmark{\PUC},
T.~M.~Crawford\altaffilmark{\KICPChicago,\AAUChicago},
A.~T.~Crites\altaffilmark{\KICPChicago,\AAUChicago},
S.~Desai\altaffilmark{\NCSA,\Illinois},
M.~A.~Dobbs\altaffilmark{\McGill},
J.~P.~Dudley\altaffilmark{\McGill},
G.~G.~Fazio\altaffilmark{\CfA},
W.~R.~Forman\altaffilmark{\CfA},
G.~Garmire\altaffilmark{\PSU},
E.~M.~George\altaffilmark{\Berkeley},
M.~D.~Gladders\altaffilmark{\KICPChicago,\AAUChicago},
A.~H.~Gonzalez\altaffilmark{\UFlorida},
N.~W.~Halverson\altaffilmark{\Colorado},
F.~W.~High\altaffilmark{\KICPChicago,\AAUChicago},
G.~P.~Holder\altaffilmark{\McGill},
W.~L.~Holzapfel\altaffilmark{\Berkeley},
S.~Hoover\altaffilmark{\KICPChicago,\EFIChicago},
J.~D.~Hrubes\altaffilmark{\UChicago},
C.~Jones\altaffilmark{\CfA},
M.~Joy\altaffilmark{\NASA},
R.~Keisler\altaffilmark{\KICPChicago,\PhysicsUChicago},
L.~Knox\altaffilmark{\Davis},
A.~T.~Lee\altaffilmark{\Berkeley,\LBNL},
E.~M.~Leitch\altaffilmark{\KICPChicago,\AAUChicago},
M.~Lueker\altaffilmark{\Berkeley},
D.~Luong-Van\altaffilmark{\UChicago},
D.~P.~Marrone\altaffilmark{\KICPChicago,\UChicago},
J.~J.~McMahon\altaffilmark{\KICPChicago,\EFIChicago,\Michigan},
J.~Mehl\altaffilmark{\KICPChicago,\AAUChicago},
S.~S.~Meyer\altaffilmark{\KICPChicago,\EFIChicago,\PhysicsUChicago,\AAUChicago},
J.~J.~Mohr\altaffilmark{\Munich,\ExcellenceCluster,\MPE},
T.~E.~Montroy\altaffilmark{\CaseWestern},
S.~S.~Murray,\altaffilmark{\CfA}
S.~Padin\altaffilmark{\KICPChicago,\AAUChicago,\Caltech},
T.~Plagge\altaffilmark{\KICPChicago,\AAUChicago},
C.~Pryke\altaffilmark{\KICPChicago,\EFIChicago,\AAUChicago},
C.~L.~Reichardt\altaffilmark{\Berkeley},
A.~Rest\altaffilmark{\Harvard,\STScI},
J.~E.~Ruhl\altaffilmark{\CaseWestern},
B.~R.~Saliwanchik\altaffilmark{\CaseWestern},
A.~Saro\altaffilmark{\Munich},
K.~K.~Schaffer\altaffilmark{\KICPChicago,\EFIChicago},
L.~Shaw\altaffilmark{\McGill,\Yale},
E.~Shirokoff\altaffilmark{\Berkeley},
J.~Song\altaffilmark{\Michigan},
H.~G.~Spieler\altaffilmark{\LBNL},
B.~Stalder\altaffilmark{\CfA},
S.~A.~Stanford\altaffilmark{\Davis},
Z.~Staniszewski\altaffilmark{\CaseWestern},
A.~A.~Stark\altaffilmark{\CfA},
K.~Story\altaffilmark{\KICPChicago,\PhysicsUChicago},
C.~W.~Stubbs\altaffilmark{\Harvard,\CfA},
K.~Vanderlinde\altaffilmark{\McGill},
J.~D.~Vieira\altaffilmark{\KICPChicago,\PhysicsUChicago,\Caltech},
A.~Vikhlinin\altaffilmark{\CfA},
R.~Williamson\altaffilmark{\KICPChicago,\AAUChicago}, and
A.~Zenteno\altaffilmark{\Munich,\ExcellenceCluster}
}

\altaffiltext{\CfA}{Harvard-Smithsonian Center for Astrophysics, 60 Garden Street, Cambridge, MA 02138}
\altaffiltext{\Clay}{Clay Fellow. Electronic address rfoley@cfa.harvard.edu .}
\altaffiltext{\Munich}{Department of Physics, Ludwig-Maximilians-Universit\"{a}t, Scheinerstr.\ 1, 81679 M\"{u}nchen, Germany}
\altaffiltext{\MIT}{MIT Kavli Institute for Astrophysics and Space Research, Massachusetts Institute of Technology, 77 Massachusetts Avenue, Cambridge, MA 02139}
\altaffiltext{\ExcellenceCluster}{Excellence Cluster Universe, Boltzmannstr.\ 2, 85748 Garching, Germany}
\altaffiltext{\McGill}{Department of Physics, McGill University, 3600 Rue University, Montreal, Quebec H3A 2T8, Canada}
\altaffiltext{\Harvard}{Department of Physics, Harvard University, 17 Oxford Street, Cambridge, MA 02138}
\altaffiltext{\Cardiff}{Department of Physics and Astronomy, Cardiff University, CF24 3YB, UK}
\altaffiltext{\UChicago}{University of Chicago, 5640 South Ellis Avenue, Chicago, IL 60637}
\altaffiltext{\NCSA}{National Center for Supercomputing Applications, University of Illinois, 1205 West Clark Street, Urbanan, IL 61801}
\altaffiltext{\KICPChicago}{Kavli Institute for Cosmological Physics, University of Chicago, 5640 South Ellis Avenue, Chicago, IL 60637}
\altaffiltext{\EFIChicago}{Enrico Fermi Institute, University of Chicago, 5640 South Ellis Avenue, Chicago, IL 60637}
\altaffiltext{\PhysicsUChicago}{Department of Physics, University of Chicago, 5640 South Ellis Avenue, Chicago, IL 60637}
\altaffiltext{\NASA}{Department of Space Science, VP62, NASA Marshall Space Flight Center, Huntsville, AL 35812}
\altaffiltext{\AAUChicago}{Department of Astronomy and Astrophysics, University of Chicago, 5640 South Ellis Avenue, Chicago, IL 60637}
\altaffiltext{\PUC}{Departamento de Astronom'a y Astrof'sica, PUC Casilla 306, Santiago 22, Chile}
\altaffiltext{\Illinois}{Department of Astronomy, University of Illinois, 1002 West Green Street, Urbana, IL 61801}
\altaffiltext{\PSU}{Department of Astronomy and Astrophysics, Pennsylvania State University, 525 Davey Lab, University Park, PA 16802}
\altaffiltext{\Berkeley}{Department of Physics, University of California, Berkeley, CA 94720}
\altaffiltext{\UFlorida}{Department of Astronomy, University of Florida, Gainesville, FL 32611}
\altaffiltext{\Colorado}{Department of Astrophysical and Planetary Sciences and Department of Physics, University of Colorado, Boulder, CO 80309}
\altaffiltext{\Davis}{Department of Physics, University of California, One Shields Avenue, Davis, CA 95616}
\altaffiltext{\LBNL}{Physics Division, Lawrence Berkeley National Laboratory, Berkeley, CA 94720}
\altaffiltext{\Michigan}{Department of Physics, University of Michigan, 450 Church Street, Ann Arbor, MI, 48109}
\altaffiltext{\MPE}{Max-Planck-Institut f\"{u}r extraterrestrische Physik, Giessenbachstr.\ 85748 Garching, Germany}
\altaffiltext{\CaseWestern}{Physics Department, Case Western Reserve University, Cleveland, OH 44106}
\altaffiltext{\Caltech}{California Institute of Technology, 1200 E. California Blvd., Pasadena, CA 91125}
\altaffiltext{\STScI}{Space Telescope Science Institute, 3700 San Martin Dr., Baltimore, MD 21218}
\altaffiltext{\Yale}{Department of Physics, Yale University, P.O.\ Box 208210, New Haven, CT 06520-8120}

\begin{abstract}
Using the South Pole Telescope (SPT), we have discovered the most
massive known galaxy cluster at $z > 1$, \cluster.  In addition to
producing a strong Sunyaev-Zel'dovich effect signal, this system is a
luminous X-ray source and its numerous constituent galaxies display
spatial and color clustering, all indicating the presence of a massive
galaxy cluster.  VLT and Magellan spectroscopy of 18 member
galaxies shows that the cluster is at \zspas.  {\it Chandra}
observations obtained through a combined HRC-ACIS GTO program reveal
an X-ray spectrum with an Fe~K line redshifted by \zx.  These
redshifts are consistent with the galaxy colors found in extensive optical,
near-infrared, and mid-infrared imaging.  \cluster\ displays extreme
X-ray properties for a cluster, having a core-excluded temperature of
\xtemp\ and a luminosity (within $r_{500}$) of \xlum.  The combined
mass estimate from measurements of the Sunyaev-Zel'dovich effect and
X-ray data is $M_{200} = \massbest$~\munit.  The discovery of such a
massive gravitationally collapsed system at high redshift provides an
interesting laboratory for galaxy formation and evolution, and is a
probe of extreme perturbations of the primordial matter density field.
We discuss the latter, determining that, under the assumption of
$\Lambda$CDM cosmology with only Gaussian perturbations, there is only
a 7\% chance of finding a galaxy cluster similar to \cluster\ in the
2500~\sqdeg SPT survey region, and that only one such galaxy cluster
is expected in the entire sky.
\end{abstract}



\setcounter{footnote}{1}

\section{Introduction}\label{s:intro}

Galaxy clusters are the most massive collapsed objects in the
Universe.  Massive clusters 
form
from the most overdense perturbations on their mass scale in the primordial
matter density field and are exceedingly rare objects.  At high redshift, corresponding to a relatively short time
after the Big Bang, they provide a powerful probe of the
initial matter density field of the Universe.   Their abundance is a particularly sensitive
probe of Gaussianity in the matter density field
\citep[e.g.,][]{Fry86, Mortonson10}, and, as a result, cluster surveys
can test various models of inflation and topological defects
\citep[e.g.,][]{Verde01}.  Observations of galaxy clusters, in which
the constituent galaxies have similar formation histories, can
strongly constrain hierarchical formation models
\citep[e.g.,][]{DeLucia07}; high-mass clusters at high redshift are
particularly interesting astronomical laboratories for galaxy
formation and evolution.

The bulk of baryonic matter in galaxy clusters is in the form of fully
ionized intracluster gas.  The free electrons in this gas emit thermal
Bremsstrahlung radiation in the X-ray, and lead to inverse Compton
scattering of the cosmic microwave background (CMB) photons. This
scattering results in a spectral distortion of the CMB observed toward
the cluster known as the thermal Sunyaev-Zel'dovich effect
\citep[SZ;][]{Sunyaev72}.  The surface brightness of the SZ effect is
determined solely by the thermal pressure of the intra-cluster medium
integrated along the line of sight and is independent of the cluster
redshift.  The total SZ signal of a cluster is a measure of its total
thermal energy and is therefore tightly correlated with the cluster
mass.  This makes the SZ effect a powerful technique for discovering
galaxy clusters at high redshift and an efficient means of finding the
highest-mass clusters \citep[e.g.,][]{Carlstrom02}.

With the goal of detecting galaxy clusters using the SZ effect, the
10-m South Pole Telescope \citep[SPT;][]{Carlstrom09} is surveying
2500~\sqdeg in three bands at millimeter wavelengths.  The SPT survey
is sensitive to galaxy clusters with a mass $\gtrsim 5 \times
10^{14}$~\msun\ at all redshifts. The first three SZ-discovered
clusters were reported by the SPT collaboration
\citep{Staniszewski09}.  The SPT collaboration recently published the
full 5-$\sigma$ cluster catalog from the first 200 \sqdeg of the
survey, observed during the 2008 observing season
\citep{Vanderlinde10}.  Most recently, a sample of the 26 most massive
galaxy clusters extracted from the full 2500~\sqdeg SPT survey has
been used to test the standard cosmological model
\citep{Williamson11}.  We refer the reader to these previous SPT
papers for a complete description of the survey strategy and goals.
The Atacama Cosmology Telescope (ACT) has also recently reported their
initial catalog of SZ-discovered clusters
\citep{Marriage10}.  

The discovery of $z \gtrsim 1$ galaxy clusters has been challenging.
Recently, several groups have detected galaxy clusters at $z \gtrsim
1.1$, with most either being found through X-ray
\citep[e.g.,][]{Rosati04, Mullis05, Stanford06, Rosati09} or {\it
Spitzer} mid-infrared (MIR) imaging \citep[e.g.,][]{Stanford05,
Brodwin06, Eisenhardt08, Muzzin09, Wilson09, Papovich10}.  Only one
SZ-detected discovered galaxy cluster has been spectroscopically
confirmed at $z > 1 $ prior to the results presented here
\citep{Brodwin10}.  In principle, high-mass clusters at high redshift
can provide evidence for an incomplete understanding of cosmology,
much like the massive clusters at $z > 0.3$ discovered in the late
1990s provided evidence against the theoretically preferred
$\Omega_{M} = 1$ cosmological model of the time
\citep[e.g.,][]{Carlberg96, Henry97, Bahcall98, Donahue98}.

During the 2009 campaign of the SPT-SZ survey, \cluster\ was detected
with a signal-to-noise ratio (S/N) of 18.5 in 150~GHz data alone
(${\rm S/N~} = 22.1$ in multi-band data) --- the largest S/N of any
cluster in the 2008--2009 SPT-SZ sample.  Initial deep optical
follow-up observations showed no obvious clustered galaxies in
\gri\!\!, but did show a strong detection of galaxies in the $z$
band.  Additional near-infrared and {\it Spitzer} photometry further
confirmed the presence of clustered galaxies.  These extremely red
galaxies are consistent with a galaxy cluster at $z > 1$.

{\it Chandra} observations of \cluster\ revealed a luminous and
extended X-ray source.  Optical spectroscopy of member galaxies shows
that the cluster is at \zsnoe.  \cluster\ is therefore not only one of
the most massive SPT clusters, it is also the highest-redshift
spectroscopically confirmed SPT cluster.  In fact, it is the most
massive cluster known at redshift $z > 1$, making \cluster\ an
extremely interesting object for detailed study.

We present our initial detection and follow-up observations of
\cluster\ in Section~\ref{s:obs}.  In Section~\ref{s:mass}, we show
that \cluster\ is a $\gtrsim 10^{15}$~\msun\ galaxy cluster at
\zsnoe.  We examine the implications of the existence of such a
massive cluster at such high redshift in Section~\ref{s:like}.  We
summarize and conclude in Section~\ref{s:conc}.  Throughout this
paper, $M_{200}$ and $M_{500}$ masses are defined as the mass enclosed
in a spherical region which has a density 200 and 500 times the mean
and critical density of the Universe, respectively.  Uncertainties are
reported for 68\% confidence intervals.


\section{Observations, Data Reduction, \& Initial Findings}\label{s:obs}

\subsection{Microwave Observations from the South Pole Telescope}

Currently the SPT has been used to complete observations of
1500~\sqdeg to full survey depth of 18~$\uk$-arcmin at 150~GHz and and
additional 1000~\sqdeg has been surveyed to a ``preview'' depth of
54~$\uk$-arcmin, where the unit $\textrm{K}$ refers to equivalent
fluctuations in the CMB temperature.  By the end of the 2011 observing
season, we expect the full 2500~\sqdeg survey region to be complete
to full depth.  Details of the survey strategy and data processing are
described in \citet{Staniszewski09}, \citet{Vanderlinde10}, and
\citet{Williamson11}.  The 2009 observations that yielded the
discovery of \cluster\ included measurements at 95, 150, and 220~GHz
to a noise level of 42, 18, and 85~$\uk$-arcmin.

\cluster\ was detected in maps of 2009 SPT data created using 
time-ordered data processing and map-making procedures equivalent to
those described in \citet{Vanderlinde10}.  Two different cluster
extraction pipelines were used to detect and characterize clusters in
these SPT maps.  Both pipelines use a multi-scale matched-filter
approach similar to the one described in \citet{Melin06}.  One
pipeline, used to extract clusters from single-band 150~GHz data, is
identical to the cluster extraction procedure described in
\citet{Vanderlinde10}.  The other pipeline, used to extract clusters
from multi-band SPT data is identical to that described in
\citet{Williamson11}.  The final SZ observable produced by both
pipelines is $\xi$, the detection significance maximized across all
filter scales.  \cluster\ was detected at $\xi = 18.5$ by the
single-band pipeline and at $\xi = 22.1$ by the multi-band pipeline.

An image of the filtered, multi-band SPT data --- using the filter
that maximized $\xi$, and centered on \cluster\ --- is shown in the
left panel of Figure \ref{f:image}.  Detection significance contours
from this same filtered map are overlaid in Figures \ref{f:image} and
\ref{f:xrayimage}.

\subsection{Optical and Infrared Imaging}

In June 2009, we obtained \griz\!\!\! imaging of \cluster\ with the
Inamori Magellan Areal Camera and Spectrograph
\citep[IMACS;][]{Dressler06} and Low Dispersion Survey Spectrograph
(LDSS3\footnote{http://www.lco.cl/telescopes-information/magellan/\\instruments-1/ldss-3-1/
.}) on the twin Magellan $6.5\,\mathrm{m}$ telescopes.  We obtained
additional $I$-band imaging with FORS2 \citep{Appenzeller98} on the
VLT $8\,\mathrm{m}$ telescope in September 2010.  See \citet{High10}
for a description of our observation strategy and reduction procedure.

\cluster\ was observed at the CTIO $4\,\mathrm{m}$ Blanco telescope
using the NEWFIRM imager \citep{Autry03} on 28 July 2010.  Data were
obtained in the $J$ and $K_{s}$ filters under photometric conditions.
All images were acquired using a 30 point random dither pattern.  At
each dither position, six frames with 10~s exposure times were
coadded, providing 60~s total exposure per position.  The dither
pattern was repeated as necessary to achieve total exposure times of
7200~s in $J$ and 4440~s in $K_{s}$.

NEWFIRM data were reduced using the FATBOY pipeline, originally
developed for the FLAMINGOS-2 instrument, and modified to work with
NEWFIRM data in support of the Infrared Bootes Imaging Survey
(Gonzalez, in prep.).  Individual processed frames are combined using
SCAMP and SWARP \citep{Bertin02}, and photometry is calibrated to
2MASS \citep{Skrutskie06}.  The final images in both filters have FWHM
of 1\arcsec\!\!.3.

Mid-infrared {\it Spitzer}/IRAC imaging was obtained in September 2009
as part of a larger program to follow up clusters identified in the
SPT survey.  The on-target observations consisted of $8 \times 100$~s
and $6 \times 30$~s dithered exposures at 3.6 and 4.5~$\mu$m,
respectively.\footnote{In this paper, we will refer to photometry in
these bands as [3.6] and [4.5], respectively.}  The deep 3.6~$\mu$m
observations are sensitive to passively evolving cluster galaxies down
to $0.1 L^{*}$ at $z = 1.5$.  The data were reduced following the
method of \citet{Ashby09}.  Briefly, we correct for column pulldown,
mosaic the individual exposures, resample to 0\arcsec\!\!.86 pixels
(half the solid angle of the native IRAC pixels), and reject cosmic
rays.

\begin{figure*}
\begin{center}
\epsscale{1.15}
\rotatebox{0}{
\plotone{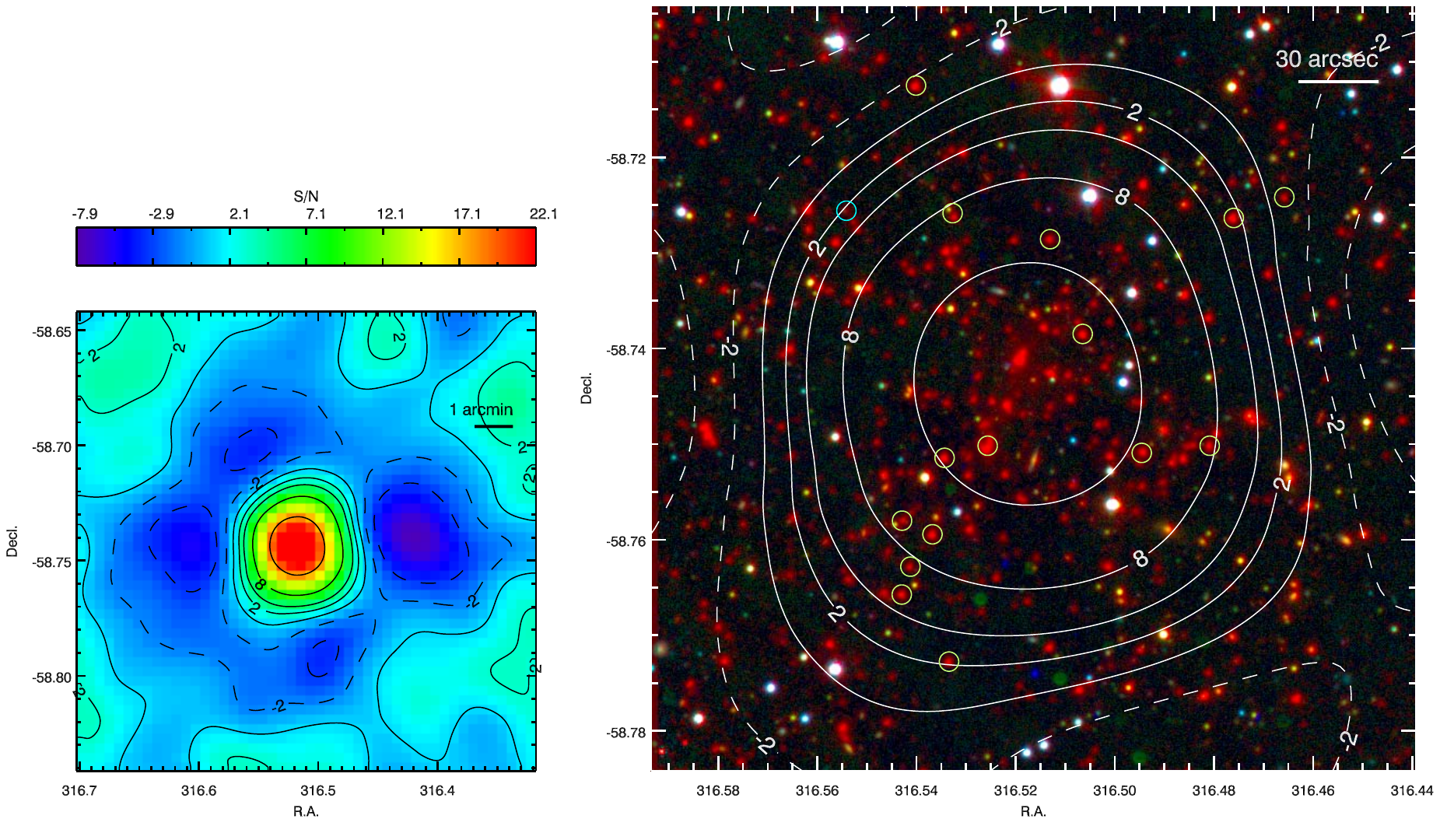}}
\caption{\cluster\ at millimeter, optical, and infrared
wavelengths.  {\it Left:} The filtered SZ significance map derived
from multi-band SPT data. The frame subtends $12\arcmin \times
12\arcmin$. The negative trough surrounding the cluster is a result of
the filtering of the time ordered data and maps. {\it Right:}
LDSS3 optical and Spitzer/IRAC mid-infrared \gione (corresponding to
BGR channels) images.  The frame subtends $4\farcm8 \times 4\farcm8$.
The white contours correspond to the SZ significance from the
left-hand panel.  The circles mark spectroscopically confirmed cluster
members, where green indicates quiescent, absorption-line member
galaxies and cyan indicates an active, emission-line member galaxy.
Some spectroscopic member galaxies are outside the FOV for this
image.}\label{f:image}
\end{center}
\end{figure*}

\begin{figure}
\begin{center}
\epsscale{1.2}
\plotone{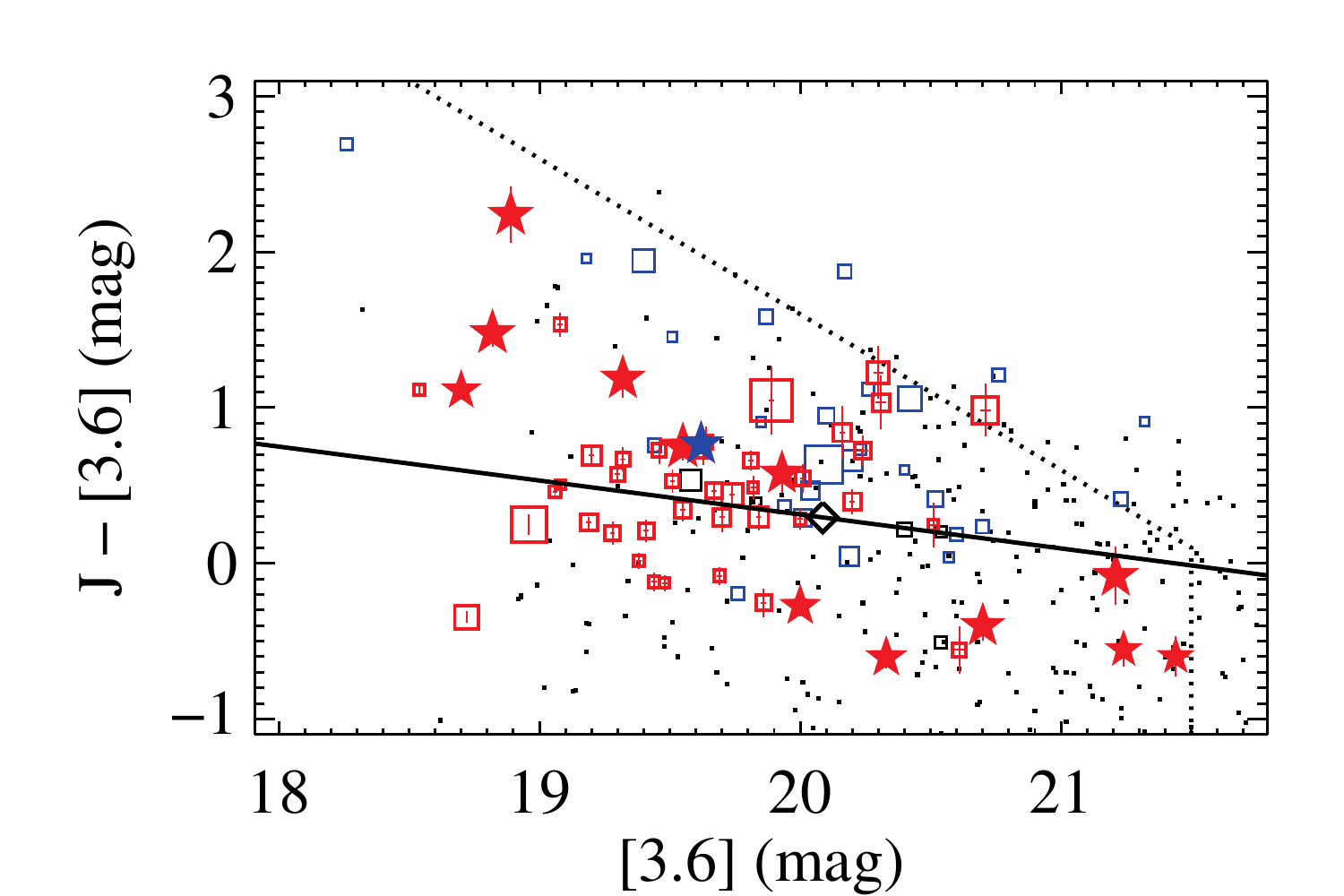}
\caption{Color-magnitude diagram ($J-[3.6]$ vs.\ [3.6]) for
galaxies within the {\it IRAC} FOV.  Suspected red-sequence cluster
members are plotted in red.  Lower-probability, but potential cluster
members are plotted in blue.  Spectroscopic members are plotted as
stars, where the red stars correspond to passive galaxies and the blue
star represents an emission-line galaxy.  Additional galaxies in the
field are plotted as black points.  The size of the symbol is
inversely proportional to the distance to the center of the cluster as
determined by the clustering of the red-sequence galaxies.  Our
5-$\sigma$ limits are plotted as dotted lines.  A red-sequence model
corresponding \zsnoe\ is represented as the solid black lines with a
representative $L^{*}$ galaxy represented by the black
diamond.}\label{f:cmd}
\end{center}
\end{figure}

Images generated from the OIR data are presented in
Figure~\ref{f:image}.  From the deep optical-only images (\gri\!\!),
it is difficult to see any indication of a cluster.  However, the
addition of infrared data shows a clear over-abundance of very red
galaxies clustered near the peak of the SZ decrement.  The
over-abundance of galaxies that are spatially coincident and have
similar colors is a clear indication of a cluster of galaxies.
Color-magnitude diagrams (CMDs) for galaxies within the imaging field
of view (FOV; Figure~\ref{f:cmd}) show an overabundance of very red
galaxies and a probable red sequence.  These data are consistent with
the spectroscopic redshifts described below.

\subsection{Optical Spectroscopy}

We obtained low-resolution spectra of several cluster members of
\cluster\ with the Gladders Image-Slicing Multislit Option
(GISMO; Gladders et~al., in prep.) module on IMACS mounted on Magellan
in June 2010; however, these spectra did not yield a definitive
cluster redshift.  Magellan slit masks were designed using galaxies
selected from deep optical and MIR imaging compatible with the red
sequence expected at $z \gtrsim 1.2$, as this was the initial redshift
approximation from the photometry.  The selection was similar to that
of \citet{Brodwin10}.  This technique gives higher priority to
brighter galaxies with colors consistent with the red sequence.  If
there is a region without a suitable red-sequence target, blue
galaxies were selected with the idea that they may be emission-line
cluster galaxies.  The Magellan observations were obtained using the
f/2 camera with the the 300 l/mm ``red'' grism and the WB6300--9500
filter.  A total of sixteen 1800~s exposures were obtained on June 6
and 7.

The VLT observations were obtained with the GRIS\_300I grism combined
with the OG590 blocking filter.  Two VLT/FORS2 masks were designed
using galaxies selected from deep $I$ band VLT pre-imaging and the
{\it Spitzer} IRAC imaging to have colors compatible with the red
sequence expected at $z \approx 1.2$ to $1.3$, which was initially
indicated by the photometry.  Each mask was populated with
approximately 30 slitlets of typical length 10\arcsec.  This target
was originally accepted as Priority B for VLT/FORS2 spectroscopy as
part of a larger SPT spectroscopy campaign.  In late November 2010,
{\it Chandra} X-ray spectroscopy gave an initial redshift from the
Fe~K line (see Section~\ref{ss:xray}).  After this result when it
became clear that there was a significant chance of it not being
observed in Fall 2010, we submitted a special request to the ESO
Director to raise this target to Priority A.  That proposal was
successful, and on each of three nights in early December 2010 a
single 2700~s exposure was obtained for a total exposure time of
8100~s on one of the two masks.  These observations were taken under
excellent seeing conditions (0\arcsec\!\!.5 to 0\arcsec\!\!.8) but at
high airmass between 1.5 and 1.7.  These data were immediately reduced
to reveal redshifts of 15 cluster members, enabling a robust cluster
redshift.

Spectroscopic reductions proceeded using standard CCD processing with
IRAF and the COSMOS reduction package \citep{Kelson03} for the VLT and
Magellan data, respectively.  The data were extracted using the
optimal algorithm of \citet{Horne86}.  Flux calibration and telluric
line removal were performed using the well-exposed continua of
spectrophotometric standard stars \citep{Wade88, Foley03}.

Three independent redshift determinations were performed using a
cross-correlation algorithm (IRAF RVSAO package; \citealt{Kurtz98}), a
template fitting method (SDSS early-type PCA templates), and a
$\chi^{2}$ minimization technique by comparing to galaxy template
spectra.  There were only minor differences in the final results from
the three methods.  In total, we have obtained secure redshifts,
consistent with membership in a single cluster, for 18 galaxies.  Two
of these galaxies have obvious [\ion{O}{2}] emission, while the others
have SEDs consistent with passive galaxies with no signs of ongoing
star formation.

\begin{figure*}
\begin{center}
\epsscale{1.1}
\rotatebox{0}{
\plotone{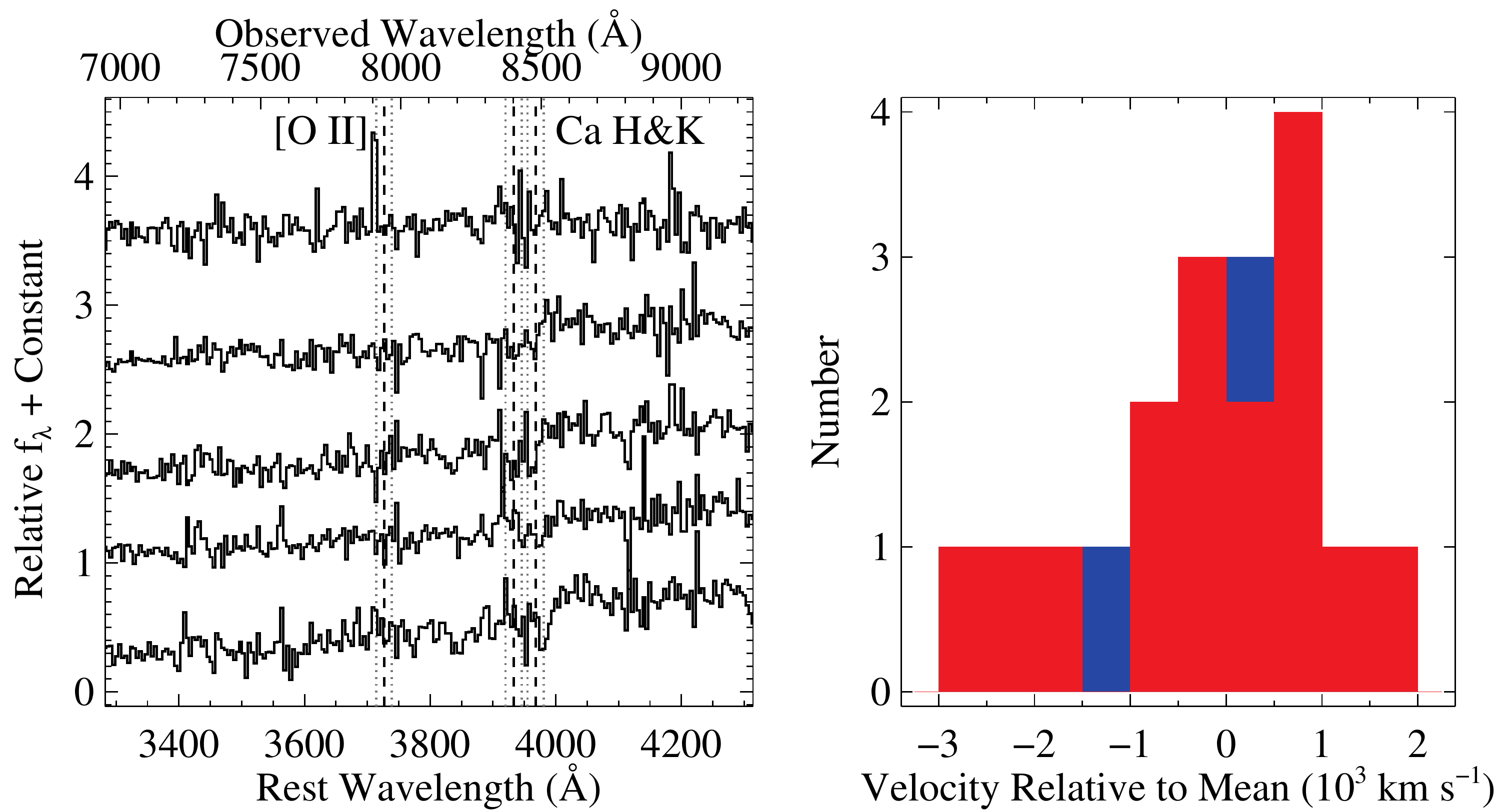}
}
\caption{{\it Left}:  Subset of optical spectra of galaxies identified
to be cluster members redshifted by the cluster redshift, \zsnoe.  The
black vertical dashed lines show the position of [\ion{O}{2}] $\lambda
3727$ and Ca H\&K $\lambda\lambda 3933$, 3968, while the grey dotted
lines show the where the spectral features would fall at $\pm
1000$~km~s$^{-1}$ relative to the cluster redshift.  {\it Right}:
Histogram of velocities for the spectroscopic cluster members relative
to the biweight mean of all members.  The blue and red histograms
represent objects with and without obvious [\ion{O}{2}] emission,
respectively.}\label{f:spec}
\end{center}
\end{figure*}

A 3-$\sigma$ clipping was applied around the peak in redshifts to
select spectroscopic cluster members.  Representative spectra of
cluster members and a redshift histogram of cluster members are
presented in Figure~\ref{f:spec}.  Redshift information for cluster
members is presented in Table~\ref{t:redshift}.  A single galaxy was
observed and has a secure redshift from both Magellan and VLT.
Although the VLT spectrum shows clear Ca H\&K absorption lines and the
Magellan spectrum only shows the D4000 break, the measured redshifts
are consistent.

A robust biweight estimator was applied to the spectroscopic sample to
determine a mean redshift of \zs\ and a velocity dispersion of \disp.
The uncertainty in both quantities is determined through bootstrap
resampling.  Since the dynamics of passive and star-forming galaxies
within clusters are different with passive galaxies assumed to be a
better tracer of the cluster potential \citep[e.g.,][]{Fadda96,
Girardi96, Mohr96, Koranyi00}, we also provide the mean redshift and
velocity dispersion of only the passive galaxies, \zspas\ and
\disppas, respectively.

The velocity distribution of spectroscopic members relative to the
mean velocity (Figure~\ref{f:spec}) is skewed.  The bootstrap
resampling produced a positive skewness in 89\% of the samples, which
indicates that it is not very statistically significant given our
sample size.  However, this may be an indication of substructure,
consistent with the possible merger seen in the X-ray image (see
Section~\ref{ss:xray}).  There is no spatial segregation of galaxies
clustered in velocity space; however, the number of spectroscopic
members is still small.

\begin{center}
\begin{deluxetable}{lllll}
\tabletypesize{\scriptsize}
\tablewidth{0pt}
\tablecaption{Redshifts for Galaxy Members of \cluster\label{t:redshift}}
\tablehead{
\colhead{R.A.\ ($^{\circ}$)\tablenotemark{a}} &
\colhead{Dec.\ ($^{\circ}$)} &
\colhead{$z$\tablenotemark{b}} &
\colhead{Feature\tablenotemark{c}} &
\colhead{Telescope}}
\startdata

316.46581                  & $-58.72418$ & 1.1365           (6) & Ca H\&K      & VLT \\
316.47601                  & $-58.72635$ & 1.1347           (3) & Ca H\&K      & VLT \\
316.48048                  & $-58.69972$ & 1.1230           (3) & [\ion{O}{2}] & VLT \\ 
316.48086                  & $-58.75017$ & 1.1288           (2) & Ca H\&K      & VLT \\
316.49456                  & $-58.75092$ & 1.1148           (2) & Ca H\&K      & VLT \\
316.50647\tablenotemark{d} & $-58.73848$ & 1.132\phantom{0} (20)  & D4000        & Magellan \\
316.50647\tablenotemark{d} & $-58.73848$ & 1.1296           (3) & Ca H\&K      & VLT \\
316.51307                  & $-58.72857$ & 1.1312           (3) & Ca H\&K      & VLT \\
316.52563                  & $-58.75017$ & 1.1450           (6) & Ca H\&K      & VLT \\
316.53273                  & $-58.72584$ & 1.120\phantom{0} (20)  & D4000        & Magellan \\
316.53346                  & $-58.77275$ & 1.1335           (5) & Ca H\&K      & VLT \\
316.53434                  & $-58.75143$ & 1.1389           (4) & Ca H\&K      & VLT \\
316.53679                  & $-58.75943$ & 1.1274           (6) & Ca H\&K      & VLT \\
316.54011                  & $-58.71250$ & 1.1384           (3) & Ca H\&K      & VLT \\
316.54123                  & $-58.76282$ & 1.1384           (5) & Ca H\&K      & VLT \\
316.54297                  & $-58.76573$ & 1.1260           (4) & Ca H\&K      & Magellan \\
316.54292                  & $-58.75800$ & 1.1363           (6) & Ca H\&K      & VLT \\
316.55414                  & $-58.72558$ & 1.1325           (3) & [\ion{O}{2}] & Magellan \\
316.74972                  & $-58.72926$ & 1.1136           (5) & Ca H\&K      & VLT
\enddata

\tablenotetext{a}{All coordinates are in J2000.}
\tablenotetext{b}{Uncertainties given in units of 0.0001.}
\tablenotetext{c}{Spectroscopic feature primarily used in determining redshift.}
\tablenotetext{d}{This is the same galaxy observed with both Magellan and VLT.}

\end{deluxetable}
\end{center}

\subsection{X-ray Observations}\label{ss:xray}

\begin{figure*}
\begin{center}
\epsscale{1.15}
\rotatebox{0}{
\plotone{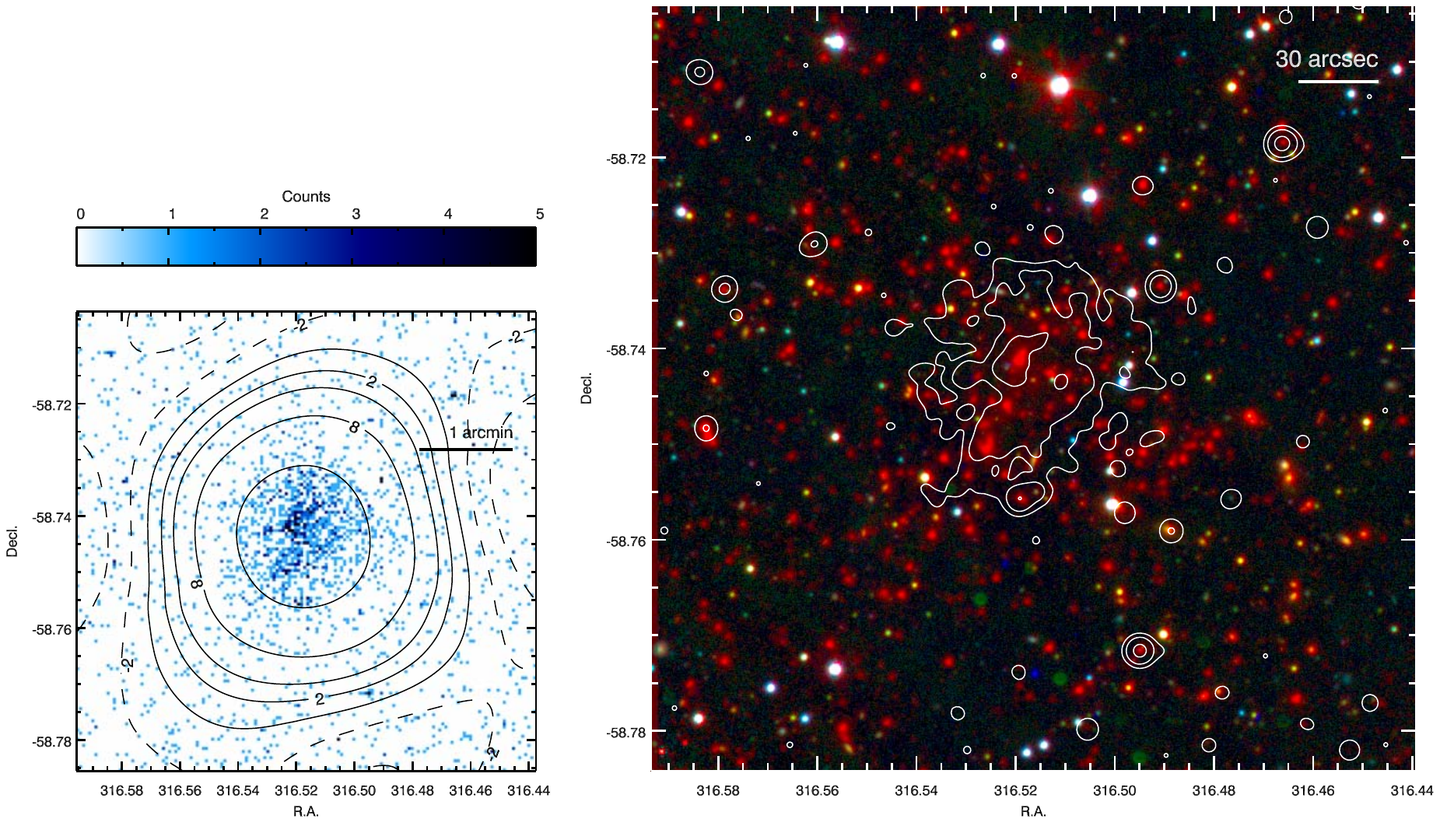}}
\caption{\cluster\ at X-ray, millimeter, optical, and infrared
wavelengths.  {\it Left:} {\it Chandra} X-ray image from the
0.7--4.0~keV band, with $4 \times 4$ binning of the original ACIS pixels
(0\farcs49 on a side).  The black contours correspond to the SZ
significance from the left-hand panel of Figure \ref{f:image}.  {\it
Right:} LDSS3 optical and Spitzer/IRAC mid-infrared \gione
(corresponding to BRG channels) images.  The frame subtends $4\farcm8
\times 4\farcm8$.  The white contours are the smoothed X-ray signal
from the left panel.}\label{f:xrayimage}
\end{center}
\end{figure*}

As part of a program to follow up the most massive, high redshift
clusters detected with the SPT in 2009, we have obtained a 25~ks X-ray
observation of \cluster.  These data were obtained through a combined
{\it Chandra} High Resolution Camera and Advanced CCD Imaging
Spectrometer Guaranteed Time Observation program.  Data reduction was
performed with the method described by \citet{Andersson10}.  The
observation resulted in 1425 source counts within $r_{500}$ in the
0.5--7.0~keV band, corresponding to a flux, $F_{X} ({\rm 0.5 -
2.0~keV}) = (2.58 \pm 0.15) \times 10^{-13}$~erg~cm$^{-2}$~s$^{-1}$.
A $2\arcsec \times 2\arcsec$ -binned X-ray image was produced using
the X-ray events in the 0.7--4.0~keV band (Figure~\ref{f:xrayimage}).
The observation reveals an X-ray luminous cluster with pronounced
asymmetry in the surface brightness distribution (see
Figure~\ref{f:xrayimage}), indicating that it may be in a merging
state.  In particular, we note that a bright concentration of gas is
offset \about 10\arcsec east of the main cluster centroid.

X-ray observations of \cluster\ were used to measure the gas
temperature, $T_{X}$, and the X-ray analog of Comptonization, $Y_{X} =
M_{\rm gas} \times T_{X}$.  The cluster mass can then be inferred with
the $M$--$Y_{X}$ and $M$--$T_{X}$ relations of \citet{Vikhlinin09}.
The $r_{500}$ radius is derived iteratively from the $M$--$Y_{X}$
scaling relation as described in \citet{Andersson10}.  See
Section~\ref{ss:xray_info} for details regarding the X-ray mass
estimates.  A spectrum is extracted from the data within the derived
$r_{500}$ radius, excluding the core within 0.15 $r_{500}$.  The
spectrum is fit, keeping the absorbing hydrogen column fixed at the
galactic value $n_{H} = 4.32 \times 10^{20}$~cm$^{-2}$, and the metal
abundance fixed at 0.3 solar.  Fixing the redshift to the mean
spectroscopic redshift, \zsnoe, results in a best fit \xtemp.  Within
$r_{500}$ (117\arcsec), including the cluster core, \cluster\ has a
luminosity of \xlum, comparable to the luminosity of the Bullet
cluster \citep{Markevitch02}.

\begin{figure*}
\begin{center}
\epsscale{1.15}
\plottwo{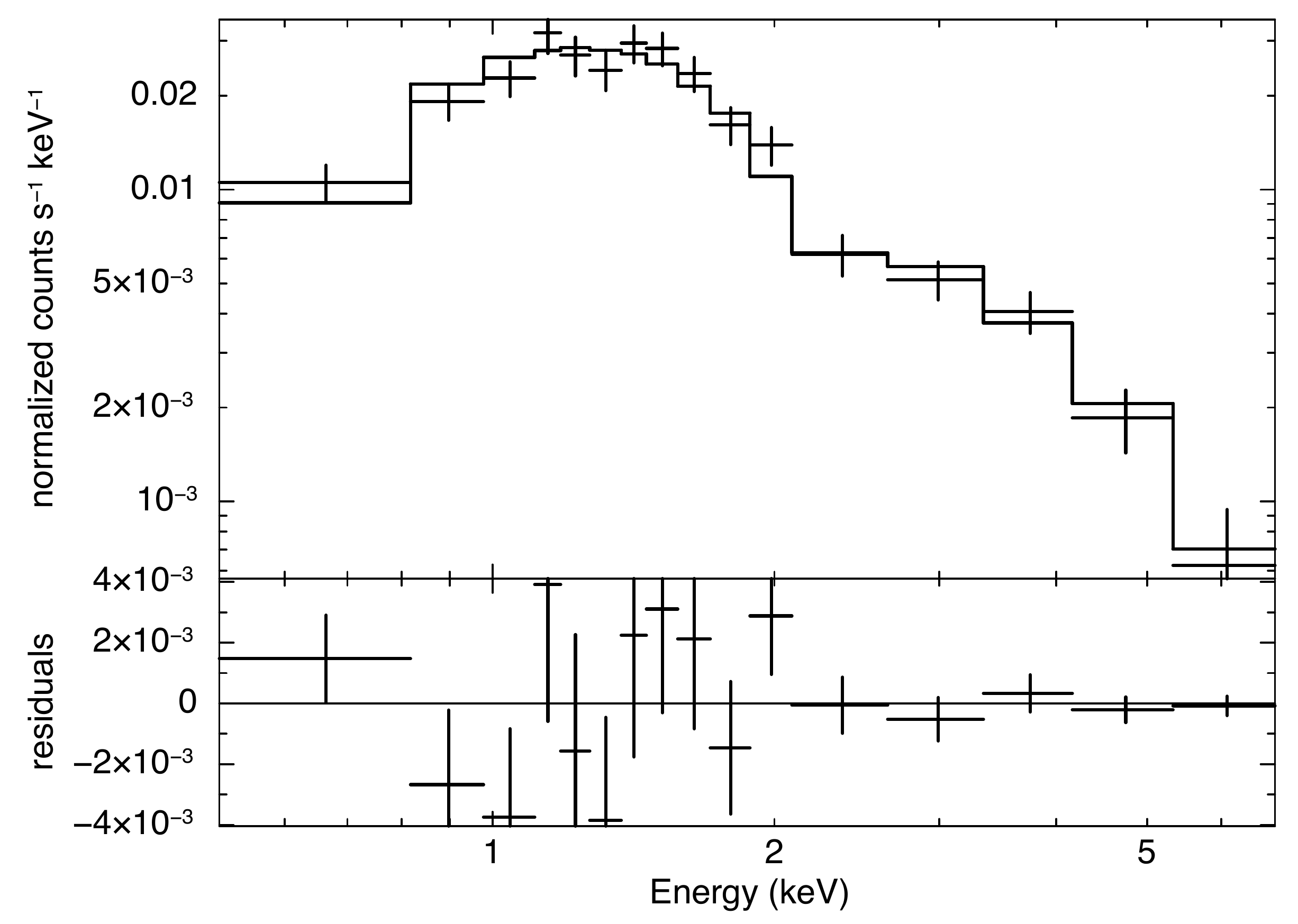}{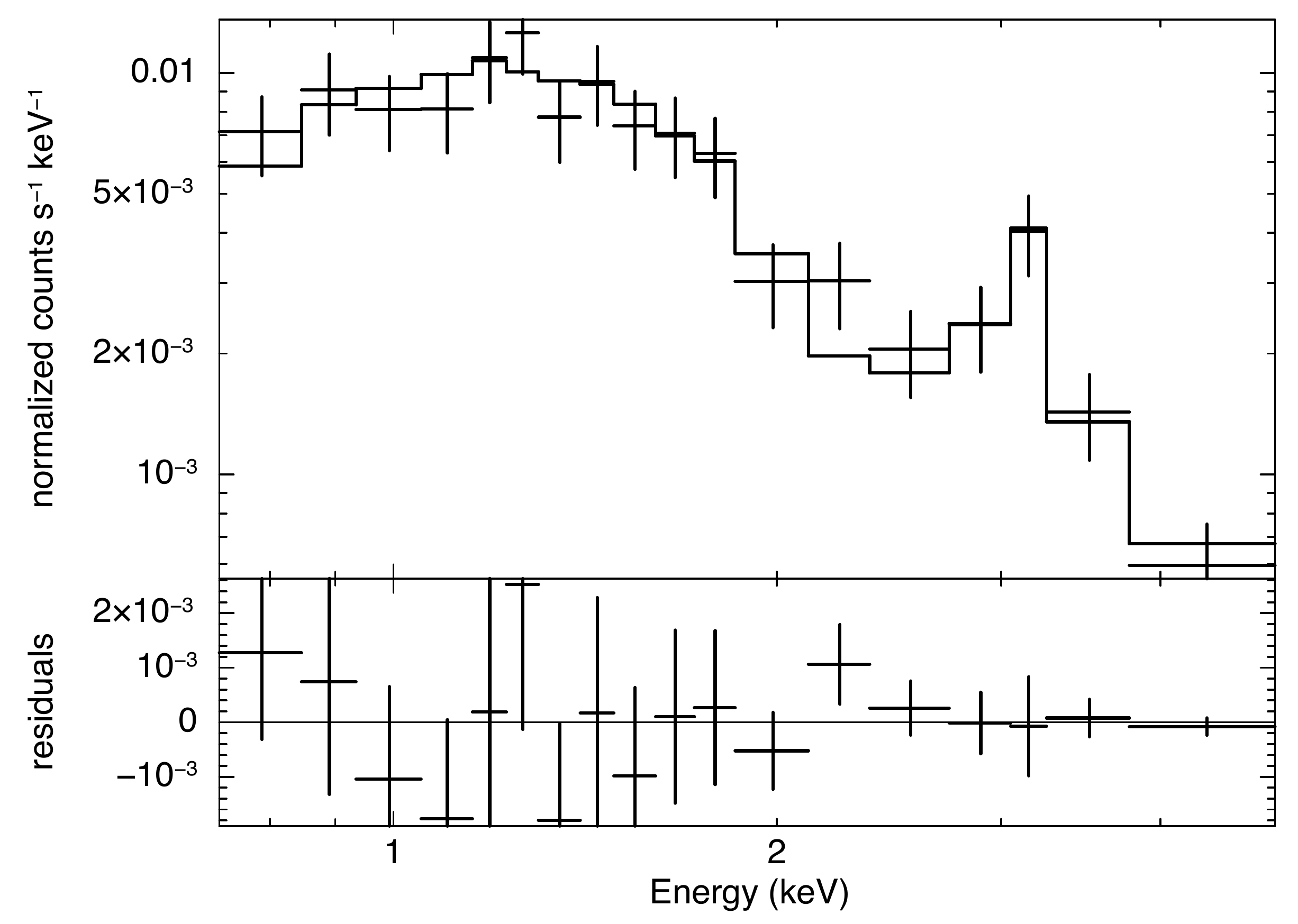}
\caption{Core-excluded (circular annulus spanning 0.15--1 $r_{500}$) and
central region (circular aperture with radius 0.17 $r_{500}$) X-ray
spectra of \cluster\ ({\it left} and {\it right} panels,
respectively).  Overplotted (solid line) is a fit APEC spectrum
redshifted to \zxnoe\ and folded through the detector sensitivity.
The 6.7~keV Fe~K line observed at \about 3~keV is both obvious and
strong for the spectrum from the central region.}\label{f:xspec}
\end{center}
\end{figure*}

An X-ray spectrum was produced from the data within 20\arcsec (0.17
$r_{500}$), where the metal abundance of the gas is likely to peak.
The resulting spectrum is shown in Figure~\ref{f:xspec} with a model
folded through the detector response.  The strong 6.7~keV Fe~K line is
present in the spectrum at approximately 3~keV.  This spectrum
indicates \zx, a metal abundance $Z = 1.44 \err{0.66}{0.51}~Z_{\sun}$,
and a central temperature $kT = 6.5 \err{1.7}{1.1}$~keV.  Notably,
this measurement was the first robust spectroscopic redshift
measurement for this cluster.

Merger simulations suggest that high-redshift cool-core clusters
should be rare \citep{Gottlober01}, and previous observations have
indicated a steep decline in the number of cool-core clusters at high
redshift \citep[e.g.,][]{Vikhlinin07}.  However, other analyses have
found that the cool-core fraction varies significantly between
different high redshift X-ray selected samples
\citep[e.g.,][]{Santos10}.  Future work comparing these results to the
cool-core fraction in SZ selected samples should help to understand
the role of selection in these results.  Regardless of the abundance
at high redshift, \cluster\ is one of the few high-redshift cool-core
galaxy clusters known.


\section{Mass Estimates from Cluster Observables}\label{s:mass}

In estimating the mass of a galaxy cluster from any observable
property, care must be taken to properly account for Eddington bias,
which, in this particular case, refers to the preferential scattering
of lower-mass systems into a given observable bin due to a steeply
falling source population.  This is handled through application of
Bayes' theorem: we use scaling relations and noise measurements to
calculate conditional probabilities of observable $O$ at fixed mass,
$P(O | M)$, and combine these with a mass prior $P(M)$ to yield a
posterior estimate of the mass $P(M | O)$.  When the observable $O$ is
the same quantity that was used to detect the cluster, the prior
probability of the cluster's mass is simply the cosmological mass
function.  However, when the observable in question is from a targeted
follow-up observation, the prior must also take into account the
selection function of the survey in which it was discovered.  For
follow-up observations targeting a single, very rare or interesting
object, such as \cluster, the appropriate prior mass distribution is
strongly influenced by the selection.  For this reason, we report
masses derived from follow-up observations with a flat mass prior
only.  We report both flat-prior and mass-function-prior masses for
the SZ-derived mass.  In Section~\ref{ss:combined_mass}, we combine
the conditional probabilities derived from different observables and
apply the mass-function prior to the result in order to determine the
best mass estimate based on all the available information.  For the
cosmological analyses in Section~\ref{s:like}, we use only this joint
posterior mass probability density.

\subsection{SZ Mass Estimate}\label{ss:sz_mass}

An SZ-based cluster mass estimate is produced following
\citet{Vanderlinde10}, using the single-band (150~GHz) detection
significance $\xi$ as a mass proxy. We direct the reader to that work
for details. Briefly, an SZ-mass scaling relation, derived from the
simulations of \citet{Shaw09} and constrained though a cosmological
analysis, is used to estimate a set of conditional probabilities
$P(\xi | M)$, taking into account the measurement noise and the
intrinsic scatter in the scaling relation.  A posterior mass estimate
is then calculated through application of a mass prior.  The
systematic errors are estimated by linearly expanding the mass
estimate as a function of the scaling relation parameters and using
the marginalized uncertainties on these parameters.

The SZ detection of this cluster is from an untargeted survey,
allowing us to take the halo mass function of \citet{Tinker08} as a
prior on its mass.  We assume the WMAP 7-year preferred $\Lambda$CDM
parameters \citep{Larson10} when calculating this prior; marginalizing
over the $\Lambda$CDM cosmological parameter uncertainties was found
to have a negligible impact on the effect of this prior.  \cluster\
has a single-band 150~GHz significance of $\xi = 18.5$, which results
in an unbiased posterior SZ-derived mass estimate of $M_{{\rm SZ},
200} = \masssz$~\munit.

We also report (see Section~\ref{ss:combined_mass} and
Table~\ref{t:mass}) a biased SZ mass estimate, using a flat mass
prior; we present it here for completeness only.  It is important to
note that this mass is not directly comparable to other
flat-mass-prior estimates (e.g., the masses estimated in
Sections~\ref{ss:dyn_mass} and \ref{ss:xray_info}).  The SZ
observation was drawn from an untargeted survey and hence from the
entire underlying population of clusters (given by the halo mass
function), while the targeted observations were conducted with {\it a
priori} knowledge that a massive cluster was present; these two
different modes of observation draw from different populations and
thus will exhibit different levels of Eddington bias.  To reiterate,
naively using any mass estimate without understanding the selection of
the observed population can result in biased measurements.

\subsection{Dynamical Mass Estimate}\label{ss:dyn_mass}

The dynamical mass of \cluster\ can be estimated from the velocity
dispersion using the $M$--$\sigma_{v}$ calibration of
\citet{Evrard08}.  Numerical simulations indicate that galaxies are
likely to exhibit velocity bias compared to the dark matter velocity
field \citep[e.g.,][]{Biviano06, White10}.  Mock observations of
simulated clusters that model the galaxy selection within followup
observations like those reported here enable these biases to be
quantified and corrected (Saro et~al., in prep).  Published results
suggest these biases are at the level of 10\% in the dispersion
\citep{White10}.  Here we attempt no correction to the dispersion but
account for this additional systematic uncertainty in the mass
estimator.

As the dynamical mass estimate arises from a targeted follow-up
observation, the underlying mass distribution can be well-defined (and
accounted for) only in the context of the SZ-based selection.  Such a
joint estimate of mass is presented below in
Section~\ref{ss:combined_mass}; here we report an independent (hence
biased) dynamical mass estimate using a flat prior on the mass.

Using the velocity dispersion from all 18 spectroscopic cluster
members gives $M_{{\rm dyn}, 200} = \massdyn$~\munit.  Using the
velocity dispersion from only the passive galaxies provides a very
similar result, $M_{{\rm dyn}, 200} = \massdynpass$~\munit).  In both
cases, the uncertainty takes into account the statistical uncertainty
of the velocity dispersion, an additional 13\% intrinsic scatter of
the velocity dispersion due to the orientation of the velocity
ellipsoid along the line of sight \citep{White10}, the 4\% intrinsic
scatter of the \citet{Evrard08} $M$--$\sigma_{v}$ relation, and a 10\%
estimate of the systematic uncertainty in the dispersion that arises
from the currently uncertain kinematic relationship between the
galaxies and the underlying dark matter.

\subsection{X-Ray Mass Estimate}\label{ss:xray_info}

As described in Section~\ref{ss:xray}, the X-ray observations of
\cluster\ were used to measure the gas temperature, $T_{X}$, and the
X-ray analog of Comptonization, $Y_{X} = M_{\rm gas} \times T_{X}$.
Following the methodology of \citet{Andersson10}, the data were
iteratively fit to determine $r_{500}$, $T_{X}$, and $Y_{X}$.  Using
the scaling relations of \citet{Vikhlinin09}, these values provide
estimates of the gas and total mass of the cluster, $M_{{\rm gas},
500} = (1.27 \pm 0.07) \times 10^{14}$~\munit\ and $M_{Y_{X},500} =
\massyxf$~\munit, respectively.  Estimating the cluster mass from the
gas temperature, we find a similarly high mass of $M_{T_{X}, 500} =
\masstxf$~\munit, where we have increased the mass by 17\% according
to the recipe in \citet{Vikhlinin09} because of the large degree of
asymmetry in the surface brightness distribution, indicative of
merging activity.

For both mass estimates above we have adopted a systematic uncertainty
of 9\% for the mass calibration of the scaling relations.  The
uncertainty on the self-similar evolution of these relations was
estimated using the simulation informed estimate of 5\% and 7\%
uncertainty on the normalization at $z = 0.6$ for the $M$--$Y_{X}$ and
$M$--$T_{X}$, respectively. This uncertainty on the evolutionary term
was extrapolated to the cluster redshift.  For further discussion on
these uncertainties, see \citet{Vikhlinin09}; the appropriate mass
prior to account for Eddington bias is only well-defined in the
context of SZ-based selection.

\subsection{Combined Mass Estimate}\label{ss:combined_mass}

The X-ray, SZ, and optical redshift data can each be used to produce
an estimate of the mass of \cluster.  We list our estimates for the
mass of \cluster\ from each of these observables in
Table~\ref{t:mass}.  Note that we have converted the $M_{Y_{X}, 500}$
mass to $M_{Y_{X}, 200}$ assuming a Navarro-Frenk-White density
profile and the mass-concentration relation of \citet{Duffy08}.
$M_{T_{X}, 500}$ is not used in the combined mass estimate since it is
correlated with $M_{Y_{X}, 500}$ and has a larger uncertainty.
Similarly, $M_{{\rm dyn}, 200}$ is not used in the combined mass
estimate because of its large uncertainty.

\begin{deluxetable}{l@{ }l@{ }l@{ }}
\tabletypesize{\scriptsize}
\tablewidth{0pt}
\tablecaption{Mass Estimates for \cluster\label{t:mass}}
\tablehead{
\colhead{Observable} &
\colhead{Measurement} &
\colhead{$M_{200} (10^{15} h_{70}^{-1} \mathrm{M}_{\sun})$}}

\startdata

{\bf SZ} $\mathbf{\xi}$       & 18.5                                     & $\mathbf{\masssztf}$ \\
SZ $\xi$ (flat prior)         & 18.5                                     & $1.24 \pm 0.30$ \\
$Y_{X}$                       & $(14.3 \pm 3.4) \times 10^{14}$ M$_{\sun}$ keV & \massyxttf \\
$T_{X}$                       & \xtempt\ keV                             & \masstxttf \\
$\sigma_{v}$\tablenotemark{a} & \disppast\ km~s$^{-1}$                   & \massdynpasstf \\
\hline\\

{\bf Combined}                & \nodata                                  & $\mathbf{\massbesttf}$

\enddata

\tablecomments{Unbolded masses indicate Eddington-biased mass
estimates, calculated using flat priors on mass.  Note that the SZ
(untargeted) measurement suffers from a considerably different
Eddington bias than the other (targeted) flat-prior estimates.  The
probability distribution for the combined mass was obtained by
multiplying the unbiased (bold) SZ mass probability by the flat-prior
$Y_{X}$ mass probability (see text for details).  Only the unbiased SZ
$\xi$ and $Y_{X}$ mass estimates were used to generate the combined
mass estimate.}
\tablenotetext{a}{As determined by the biweight estimate of the
passive galaxies.}
\end{deluxetable}

We assume for the purposes of the combined mass estimate that the
uncertainties are uncorrelated between the two masses.  In principle,
the different observables could have correlated scatter in their
scaling with mass, however, this is expected to be negligible at the
current S/N of the observables.  This simplification allows us to
write the posterior mass distribution as a product of the two
independent conditional probabilities,
\begin{equation}
  P(M|\xi, Y_{X}) \propto P(M) \cdot P(\xi | M) \cdot P(Y_{X} | M),
\end{equation}
where $P(M)$ is again the halo mass function of \citet{Tinker08},
evaluated at the WMAP 7-year preferred point in $\Lambda$CDM parameter
space.  Functionally, this reduces to a product of two probability
density functions, the unbiased SZ mass estimate and the flat-prior
$Y_{X}$ mass estimate.  We find the combined, unbiased, mass estimate
from the two observables to be $M_{200} = \massbest$~\munit.

All mass estimates are consistent and very high for a cluster at this
redshift.  Each individual mass estimate, as well as the combined
estimate, indicates that \cluster\ is the most massive cluster yet
discovered at $z > 1$.  It is \about 60\% more massive than
SPT-CL~J0546-5345 \citep{Brodwin10} at $z = 1.07$, the second most
massive galaxy cluster at $z > 1$.


\section{Likelihood Analysis}\label{s:like}

Since \cluster\ is the highest S/N cluster found in 2008-2009 SPT
survey fields, the highest-redshift spectroscopically confirmed SPT
cluster, and the most massive cluster discovered at $z > 1$ by any
method, it raises the question: Is the existence of \cluster\
consistent with a Gaussian-distributed primordial density field in a
$\Lambda$CDM universe?

We investigate the probability of finding a cluster with the
mass and redshift of \cluster\ in the context of $\Lambda$CDM as follows.
Based on the likelihood calculation of \citet{Vanderlinde10} and the
CosmoMC software package \citep{Lewis02}, we produce an MCMC chain
sampling the parameter space of spatially flat $\Lambda$CDM cosmology,
storing the halo mass function of \citet{Tinker08} at each step in the
chain.  Current data sets, consisting of the WMAP 7-year data
\citep{Larson10} and the SPT 2008 cluster catalog
\citep{Vanderlinde10}, are used to constrain the parameter space
explored.  This yields a set of halo mass functions representative of
current uncertainties of $\Lambda$CDM parameter space.

Integrating each of these mass functions, we compute the Poisson
likelihood of finding a cluster at or above the mass and redshift of
\cluster\ in a 2500~\sqdeg field.  For the MCMC chain described above,
the existence of a cluster at or above the mass and redshift of
\cluster\ is less than 5\% likely in models contributing 32\% of the
WMAP7+V10 probability.  Marginalized over all WMAP7+V10 allowed
parameter space, we find that there is a 7\% chance of finding a
cluster at least as massive and at a redshift at least as high as
\cluster\ in the SPT survey field.  Prior to having observed this
cluster, the expectation would be that there is only one galaxy
cluster at least as massive and at a redshift at least as high as
\cluster\ in the entire sky.


\section{Conclusions}\label{s:conc}

We present the discovery and initial characterization of \cluster.
This massive cluster was discovered in the 2500~\sqdeg SPT survey as a
highly significant SZ decrement.  Initial follow-up in the optical and
infrared indicated that it was both at high redshift and very rich.
Optical spectroscopy yielded a cluster redshift of \zsnoe, consistent
with the redshift found from the X-ray spectrum of the hot
intracluster gas, \zxnoe, and the broad-band SEDs of member galaxies.
\cluster\ is currently the highest-redshift SZ-discovered galaxy
cluster.

\cluster\ is also extremely massive.  X-ray, SZ, and velocity
dispersion data provide complementary and consistent measurements of
the mass (see Table~\ref{t:mass}).  The best combined estimate of the
mass is $M_{200} = \massbest$~\munit, making \cluster\ the most
massive galaxy cluster yet discovered at $z > 1$.  The combination of
its high mass and high redshift make \cluster\ a very interesting
object in which to study galaxy formation and evolution only a few Gyr
after the formation of the first stars.  The X-ray data show that
\cluster\ has a cool core and may be currently undergoing a merger.
The distribution of galaxy velocities is slightly skewed, but is
consistent with a Gaussian distribution.  Additional spectroscopy may
help determine if there is significant substructure in the cluster.

The high mass and redshift of \cluster\ also make it an interesting
object from a cosmological perspective.  The chances of finding a
cluster this extreme in mass and redshift in a 2500~\sqdeg survey is
only 7\%, given current cosmological parameter constraints, with only
one such cluster expected in the entire sky.  As the SPT cluster
catalog grows, and our understanding of the various mass-observable
scaling relations improves, SZ-discovered clusters will enable yet
more powerful tests of the current cosmological paradigm.

\begin{acknowledgments} 

{\it Facilities:} \facility{Blanco (NEWFIRM)}, \facility{CXO (ACIS)},
\facility{Magellan:Baade (IMACS)}, \facility{Magellan:Clay (LDSS3)},
\facility{Spitzer (IRAC)}, \facility{South Pole Telescope},
\facility{VLT:Antu (FORS2)}

\bigskip

The South Pole Telescope program is supported by the National Science
Foundation through grant ANT-0638937.  Partial support is also
provided by the NSF Physics Frontier Center grant PHY-0114422 to the
Kavli Institute of Cosmological Physics at the University of Chicago,
the Kavli Foundation, and the Gordon and Betty Moore Foundation.  This
work is based in part on observations obtained with the Spitzer Space
Telescope (PID 60099), which is operated by the Jet Propulsion
Laboratory, California Institute of Technology under a contract with
NASA.  Support for this work was provided by NASA through an award
issued by JPL/Caltech.  Additional data were obtained with the 6.5~m
Magellan Telescopes located at the Las Campanas Observatory,
Chile. Support for X-ray analysis was provided by NASA through Chandra
Project Numbers 12800071 and 12800088 issued by the Chandra X-ray
Observatory Center, which is operated by the Smithsonian Astrophysical
Observatory for and on behalf of NASA under contract NAS8-03060.
Observations from VLT programs 086.A-0741 and 286.A-5021 were included
in this work.  We acknowledge the use of the Legacy Archive for
Microwave Background Data Analysis (LAMBDA).  Support for LAMBDA is
provided by the NASA Office of Space Science.  Galaxy cluster research
at Harvard is supported by NSF grant AST-1009012.  Galaxy cluster
research at SAO is supported in part by NSF grants AST-1009649 and
MRI-0723073.  The McGill group acknowledges funding from the National
Sciences and Engineering Research Council of Canada, Canada Research
Chairs program, and the Canadian Institute for Advanced Research.
X-ray research at the CfA is supported through NASA Contract NAS
8-03060.  The Munich group acknowledges support from the Excellence
Cluster Universe and the DFG research program TR33.  R.J.F.\ is
supported by a Clay Fellowship.  B.A.B\ is supported by a KICP
Fellowship, support for M.Brodwin was provided by the W. M. Keck
Foundation, M.Bautz acknowledges support from contract
2834-MIT-SAO-4018 from the Pennsylvania State University to the
Massachusetts Institute of Technology.  M.D.\ acknowledges support
from an Alfred P.\ Sloan Research Fellowship, W.F.\ and C.J.\
acknowledge support from the Smithsonian Institution, and B.S.\
acknowledges support from the Brinson Foundation.

\end{acknowledgments}

\bibliographystyle{fapj}
\bibliography{spt}


\end{document}